\documentclass[aps,pre,twocolumn,superscriptaddress,floatfix,nofootinbib]{revtex4-2}

\usepackage{amsmath,amssymb}
\usepackage{graphicx}
\usepackage{dcolumn}
\usepackage{siunitx}
\usepackage{bm}
\usepackage[colorlinks=true,allcolors=blue]{hyperref}

\newcolumntype{d}{D{.}{.}{-1}}  % decimal-aligned columns for tables
\newcommand{\Bt}{\mathrm{B}}

\begin{document}

\title{The ring wants to be broken}

\author{Alexei Vazquez}
\email{alexei.vazquez@gmail.com}
\affiliation{Nodes \& Links Ltd, Salisbury House, Station Road,
Cambridge, CB1 2LA, UK}

\date{\today}

\begin{abstract}
The Ramsey community number $r_\kappa$ is the minimum network size at which
a graph's connectivity is better described by a partition into communities
than by no partition, under a prescribed community-detection rule. It was
introduced through numerical simulations of networks grown by local rules,
which suggested that community structure can emerge without any node
heterogeneity. Here I compute $r_\kappa$ \emph{analytically} for the simplest
homogeneous, locally wired graph: the circulant ring lattice $C_n(1,\dots,c)$.
Using a Bernoulli stochastic block model with symmetric $\mathrm{Beta}$ priors
as the detection rule, the Bayesian evidence for a balanced two-community
partition and for the unpartitioned network are both obtained in closed form,
so the transition between them can be located exactly. The result is a sharp
dependence on the interaction range: the plain cycle ($c=1$) is never
partitioned, its two-community posterior decaying as $n^{-(2\alpha+3)}$, so
$r_\kappa=\infty$; but the next-nearest-neighbour ring ($c=2$) acquires a
finite $r_\kappa\simeq 35$ nodes, above which the partition is preferred with
a log-evidence growing as $(\ln 2)\,n$. This provides an exactly solvable
instance of community emergence in a network with no built-in communities,
and shows that a minimal amount of local connectivity is enough to break the
ring.
\end{abstract}

\maketitle

\begin{figure*}[t]
\centering
\includegraphics[width=\textwidth]{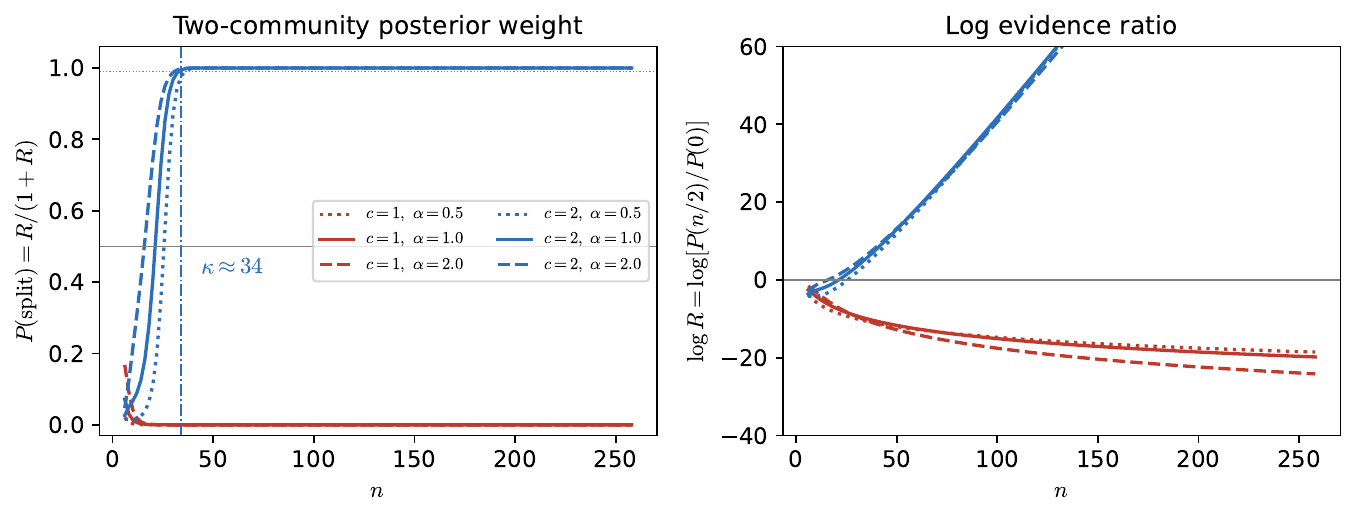}
\caption{(a) Posterior weight of the balanced two-community partition,
Eq.~\eqref{eq:psplit}; the dash-dotted line marks the Ramsey community number
$r_\kappa$ ($99\%$ certainty) for $c{=}2,\ \alpha{=}1$. (b) The log evidence
ratio $\log R$. Red curves are $c{=}1$, blue curves $c{=}2$; line styles denote
$\alpha=0.5$ (dotted), $1$ (solid), and $2$ (dashed). For $c=1$ the partition
is never preferred; for $c=2$ it takes over near $n\approx 30$ and
$P(\mathrm{split})\to1$.}
\label{fig:main}
\end{figure*}

% ======================================================================
\section{Introduction}
% ======================================================================

Networks are a standard language for natural and social systems, and one of
their most studied features is community structure: the organisation of nodes
into groups that are more densely connected internally than to the rest of the
network~\cite{fortunato2010community,fortunato2016community}. Detecting
communities is now a mature field, with methods ranging from the
edge-betweenness algorithm of Girvan and Newman~\cite{girvan2002community,
newman2004finding} and modularity maximisation~\cite{newman2006modularity} to
flow-based approaches such as Infomap~\cite{rosvall2008maps}. A principled
strand of this literature is built on the stochastic block model (SBM), a
generative model in which the probability of a link depends only on the block
memberships of its endpoints~\cite{holland1983stochastic}. The SBM and its
degree-corrected variant~\cite{karrer2011stochastic} turn community detection
into statistical inference: the connection probabilities and the node labels
are fitted to the observed graph, detectability is governed by a phase
transition~\cite{decelle2011asymptotic}, and the number of communities can be
selected by Bayesian model comparison~\cite{peixoto2014hierarchical,
abbe2018community}. Model selection is central here, because it is what allows
one to ask not merely \emph{how} to partition a network but \emph{whether} it
should be partitioned at all.

Community structure is usually attributed to node heterogeneity---attributes
such as political affiliation in social networks or biological function in
molecular networks that make some node pairs intrinsically more likely to
connect. Recently, however, I showed by numerical simulation that node
heterogeneity is not a necessary
condition~\cite{vazquez2025ramsey}. Networks grown by purely local
rules~\cite{vazquez2003growing}---rules in which new links attach in the
neighbourhood of existing ones, with no reference to any node label---routinely
acquire community structure as they grow. To make this quantitative, that work
introduced the \emph{Ramsey community number} $r_\kappa$: the minimum graph
size that guarantees, with near-certainty, that a prescribed detection method
(the degree-corrected SBM or regularised Infomap) reports two or more
communities. The name evokes Ramsey theory, in which a large enough structure
unavoidably contains ordered substructures~\cite{ramsey1930problem}: here a
large enough locally grown network unavoidably ``contains'' communities.
Networks generated by local rules were found to have finite $r_\kappa$, whereas
their degree-preserving randomisations did not have the emergent-community
property, leading to the conjecture that communities are an emergent property
of local network evolution~\cite{vazquez2025ramsey}.

Because $r_\kappa$ was established numerically, on stochastic ensembles and with
detection methods that must themselves be run algorithmically, it is natural to
ask whether the same quantity can be obtained analytically in a controlled
case. That is the goal of this paper. I take the most elementary homogeneous,
locally wired graph---the circulant ring lattice $C_n(1,\dots,c)$, in which
every node is joined to its $c$ nearest neighbours on each side---and use a
Bernoulli SBM with symmetric $\mathrm{Beta}$ priors as the detection rule. For
this graph the Bayesian evidence of a balanced two-community partition and of
the unpartitioned network are both single, closed-form expressions, so their
comparison, and hence $r_\kappa$, can be computed exactly rather than sampled.
The ring is deliberately featureless: every node is identical, there is no
planted structure, and the only control parameter is the interaction range
$c$. It is therefore a clean test bed for the emergence question.

The answer is a sharp threshold that depends on $c$. The plain cycle
($c=1$) is never preferentially partitioned---its two-community posterior
decays as a power of $n$---so its Ramsey community number is infinite. Adding a
single extra neighbour ($c=2$) changes the picture qualitatively: beyond a
finite size the partitioned description wins, and does so exponentially fast in
$n$, giving a finite $r_\kappa$ of order a few tens of nodes. A minimal amount
of local connectivity is thus sufficient to make the ring ``want to be
broken,'' an analytically exact example of the phenomenon reported
in~\cite{vazquez2025ramsey}. The rest of the paper sets up the model
(Sec.~\ref{sec:model}), derives the closed-form evidence ratio
(Sec.~\ref{sec:ratio}), analyses the size-driven transition
(Sec.~\ref{sec:transition}), reports the Ramsey community number
(Sec.~\ref{sec:kappa}), and closes with conclusions
(Sec.~\ref{sec:conclusions}).

% ======================================================================
\section{Model and detection rule}
\label{sec:model}
% ======================================================================

\subsection{The circulant ring}

The network is the circulant ring lattice $C_n(1,\dots,c)$: the $n$ nodes are
placed on a circle and each node $i$ is joined to its $c$ forward neighbours
$\{i{+}1,\dots,i{+}c\}\pmod n$, and hence also to its $c$ backward neighbours.
Every node has degree $2c$ and the graph has $cn$ edges. I take $n$ even,
$c\in\{1,2\}$, and work in the regime $n\gg c$. This is the base lattice of the
Watts--Strogatz small-world construction~\cite{watts1998collective}, here left
unrewired so that the graph is perfectly homogeneous.

A candidate partition cuts the ring into two contiguous arcs
$A=\{1,\dots,n_1\}$ and $B=\{n_1{+}1,\dots,n\}$, with $n_2=n-n_1$. The value
$n_1=0$ denotes \emph{no partition}, i.e. the whole ring as a single block, and
$n_1=n/2$ the balanced \emph{two-community} partition.

\subsection{Stochastic block model evidence}

Under a Bernoulli SBM with within/within/between connection probabilities
$\theta_1,\theta_2,\theta_{12}$ and a Bernoulli($\pi$) label prior, the
marginal likelihood (evidence) of a split at $n_1$ is
\begin{align}
P(n_1)=\frac{1}{Z}\int
&\prod_{i\in\{1,2\}}\theta_i^{\,E_i}(1-\theta_i)^{\binom{n_i}{2}-E_i}
\nonumber\\
&\times\,\theta_{12}^{\,E_{12}}(1-\theta_{12})^{n_1 n_2-E_{12}}
\nonumber\\
&\times\,\pi^{n_1}(1-\pi)^{\,n-n_1}\;P(\theta)\,P(\pi)\,d\theta\,d\pi ,
\label{eq:model}
\end{align}
where $E_i$ counts the edges present inside block $i$ and $E_{12}$ the edges
crossing between the two blocks. All priors are symmetric Beta with a common
hyperparameter,
\begin{align}
P(\theta_j)&=\mathrm{Beta}(\theta_j;\alpha,\alpha)\quad (j\in\{1,2,12\}),\\
P(\pi)&=\mathrm{Beta}(\pi;\alpha,\alpha).
\end{align}
The detection rule is Bayesian model comparison: prefer whichever hypothesis,
$n_1=n/2$ or $n_1=0$, has the larger evidence.

\subsection{Edge counts}

A ring has two seams, so a contiguous cut severs edges in two places and the
crossing count is a small constant, independent of $n_1$, provided the seams do
not interact ($n_1,n_2>c$). Counting edges of each length $k=1,\dots,c$ gives,
for an interior arc block of size $n_i$,
\begin{align}
E_i&=\sum_{k=1}^{c}(n_i-k)=c\,n_i-\frac{c(c+1)}{2},\label{eq:Ei}\\
\binom{n_i}{2}-E_i&=\binom{n_i-c}{2},\label{eq:Eic}
\end{align}
the second identity following because
$\tfrac12[n_i^2-(2c{+}1)n_i+c(c{+}1)]=\tfrac12(n_i-c)(n_i-c-1)$. For the
crossing block, edges are cut at both seams,
\begin{equation}
E_{12}=2\sum_{k=1}^{c}k=c(c+1),
\label{eq:E12}
\end{equation}
out of $n_1 n_2$ possible. For the unpartitioned network ($n_1=0$) the single
block is the \emph{closed cycle}, which regains the seam edges, so $E=cn$ out of
$\binom{n}{2}$. These counts satisfy edge conservation,
$E_1+E_2+E_{12}=cn$. Concretely $c=1$ gives $E_i=n_i-1,\ E_{12}=2$, and $c=2$
gives $E_i=2n_i-3,\ E_{12}=6$.

% ======================================================================
\section{Closed-form evidence ratio}
\label{sec:ratio}
% ======================================================================

With symmetric $\mathrm{Beta}(\alpha,\alpha)$ priors, each parameter integral
in Eq.~\eqref{eq:model} is one dimensional and collapses to a Beta function
through
$\int_0^1 x^{E+\alpha-1}(1-x)^{N-E+\alpha-1}dx=\Bt(E{+}\alpha,\,N{-}E{+}\alpha)$.
A block that carries no data ($E=0$ out of $N=0$ possible edges) integrates to
$1$, leaving a bare prior normaliser $\Bt(\alpha,\alpha)$ that does \emph{not}
cancel; this is a Bayesian Occam penalty for unused parameters. Writing
$m\equiv n/2$, the two evidences are
\begin{widetext}
\begin{align}
P(n/2)&=\frac{1}{Z\,\Bt(\alpha,\alpha)^4}\,
\Bt\!\Big(cm{-}\tfrac{c(c+1)}{2}{+}\alpha,\ \tbinom{m-c}{2}{+}\alpha\Big)^{\!2}\,
\Bt\!\big(c(c{+}1){+}\alpha,\ m^2{-}c(c{+}1){+}\alpha\big)\,
\Bt\!\big(m{+}\alpha,\ m{+}\alpha\big),
\label{eq:Phalf}\\[3pt]
P(0)&=\frac{1}{Z\,\Bt(\alpha,\alpha)^4}\,
\underbrace{\Bt(\alpha,\alpha)^2}_{\text{Occam factor}}\,
\Bt\!\big(cn{+}\alpha,\ \tbinom{n}{2}{-}cn{+}\alpha\big)\,
\Bt\!\big(\alpha,\ n{+}\alpha\big).
\label{eq:Pzero}
\end{align}
\end{widetext}
The global constant $Z$ and the prefactor $\Bt(\alpha,\alpha)^{-4}$ cancel in
the ratio $R\equiv P(n/2)/P(0)$,
\begin{widetext}
\begin{equation}
R=\frac{
\Bt\!\big(cm{-}\tfrac{c(c+1)}{2}{+}\alpha,\ \binom{m-c}{2}{+}\alpha\big)^{2}\,
\Bt\!\big(c(c{+}1){+}\alpha,\ m^2{-}c(c{+}1){+}\alpha\big)\,
\Bt(m{+}\alpha,\ m{+}\alpha)}
{\Bt(\alpha,\alpha)^{2}\,
\Bt\!\big(cn{+}\alpha,\ \binom{n}{2}{-}cn{+}\alpha\big)\,
\Bt(\alpha,\ n{+}\alpha)}.
\label{eq:R}
\end{equation}
\end{widetext}
The detection rule prefers the two-community partition exactly when $R>1$. The
posterior weight it assigns to that partition, restricted to these two
hypotheses, is
\begin{equation}
P(\mathrm{split})=\frac{P(n/2)}{P(n/2)+P(0)}=\frac{R}{1+R}
=\frac{1}{1+R^{-1}}.
\label{eq:psplit}
\end{equation}
In practice Eq.~\eqref{eq:R} is evaluated in logarithmic form,
$\log R=\log P(n/2)-\log P(0)$ with
$\log\Bt(x,y)=\log\Gamma(x)+\log\Gamma(y)-\log\Gamma(x{+}y)$, so that
$P(\mathrm{split})=1/(1+e^{-\log R})$; the results below use \num{80}-digit
arithmetic.

% ======================================================================
\section{The size-driven transition}
\label{sec:transition}
% ======================================================================

Equations~\eqref{eq:R}--\eqref{eq:psplit} exhibit opposite behaviour for the
two connectivities, shown in Fig.~\ref{fig:main} and Table~\ref{tab:main}. A
positive $\log R$ means the ring prefers to split.

\begin{table}[b]
\centering
\begin{ruledtabular}
\begin{tabular}{r d d c d d}
 & \multicolumn{2}{c}{$c=1$ (cycle)} & & \multicolumn{2}{c}{$c=2$} \\
\cline{2-3}\cline{5-6}
\multicolumn{1}{c}{$n$} & \multicolumn{1}{c}{$\log R$} &
\multicolumn{1}{c}{$P(\mathrm{split})$} & &
\multicolumn{1}{c}{$\log R$} & \multicolumn{1}{c}{$P(\mathrm{split})$}\\
\hline
16   & -6.272  & 1.9\times10^{-3}  & & -1.495 & 0.183 \\
32   & -9.510  & 7.4\times10^{-5}  & & 4.240  & 0.986 \\
64   & -12.881 & 2.5\times10^{-6}  & & 20.60  & 1.000 \\
128  & -16.304 & 8.3\times10^{-8}  & & 58.92  & 1.000 \\
1024 & -26.666 & 2.6\times10^{-12} & & 661.4  & 1.000 \\
\end{tabular}
\end{ruledtabular}
\caption{Log evidence ratio and posterior split weight at $\alpha=1$. For $c=1$
the partition is never preferred; for $c=2$ it takes over near $n\approx30$.}
\label{tab:main}
\end{table}

\paragraph*{Plain cycle ($c=1$): no emergence.}
Here $\log R\sim A-(2\alpha+3)\log n\to-\infty$, so
$P(\mathrm{split})\sim n^{-(2\alpha+3)}\to0$. The decay exponent is confirmed
numerically to three decimals: measured values are $3.997$, $4.993$, $6.991$,
and $8.997$ for $\alpha=0.5,1,2,3$, matching $2\alpha+3$. A plain cycle offers
no density contrast to exploit: isolating the two crossing edges cannot outweigh
the balanced-label prior together with the Occam cost of the two extra
parameters in Eq.~\eqref{eq:Pzero}. Stronger priors (larger $\alpha$) suppress
the split faster.

\paragraph*{Next-nearest-neighbour ring ($c=2$): sharp emergence.}
Here $\log R\sim(\ln 2)\,n\to+\infty$, so $P(\mathrm{split})\to1$ exponentially
fast once the transition is crossed. The slope converges to $\ln 2=0.693147$
(measured $0.693100$ at $n\sim10^{5}$) and is independent of the prior
$\alpha$. Physically, splitting the $c=2$ ring concentrates the edge density
inside the two arcs while the crossing density falls as $c(c+1)/m^2\to0$, and
this contrast is now large enough to overcome the same prior and Occam costs
that defeat the $c=1$ case.

% ======================================================================
\section{The Ramsey community number}
\label{sec:kappa}
% ======================================================================

Following~\cite{vazquez2025ramsey}, define the Ramsey community number as the
minimum network size at which the connectivity is preferentially described by
communities to a prescribed certainty $q$,
\begin{equation}
r_\kappa(c,\alpha;q)=\min\big\{\,n\ \text{even}:\ P(\mathrm{split})\ge q\,\big\},
\end{equation}
with $r_\kappa=\infty$ when no such $n$ exists. The choice $q=\tfrac12$ locates
the crossover $R=1$, where communities first become the \emph{more consistent}
description; larger $q$ enforces the ``almost certainly'' sense
of~\cite{vazquez2025ramsey}. Because $\log R$ grows linearly in $n$ for $c=2$,
the certainty rises so steeply that the value of $r_\kappa$ shifts by only a few
nodes as $q$ is tightened (Table~\ref{tab:kappa}).

\begin{table}[t]
\centering
\begin{ruledtabular}
\begin{tabular}{c c c c c c c}
$c$ & $\alpha$ & $r_\kappa^{0.5}$ & $r_\kappa^{0.90}$ & $r_\kappa^{0.99}$
& $r_\kappa^{0.999}$ & $r_\kappa^{0.9999}$\\
\hline
1 & any & $\infty$ & $\infty$ & $\infty$ & $\infty$ & $\infty$\\
\hline
2 & 0.3 & 30 & 34 & 40 & 44 & 48\\
2 & 0.5 & 26 & 32 & 36 & 42 & 46\\
2 & 1.0 & 22 & 28 & 34 & 38 & 44\\
2 & 2.0 & 16 & 24 & 32 & 38 & 42\\
2 & 3.0 & 14 & 24 & 32 & 38 & 44\\
\end{tabular}
\end{ruledtabular}
\caption{Ramsey community number $r_\kappa^{q}$ (smallest even $n$ with
$P(\mathrm{split})\ge q$). The cycle has $r_\kappa=\infty$ at every certainty
level; the $c=2$ ring has finite $r_\kappa$. At the ``almost certain'' levels
$r_\kappa\simeq32$--$40$ nodes and becomes nearly independent of the prior
$\alpha$, because the $\alpha$-free asymptotic slope $\ln 2$ dominates.}
\label{tab:kappa}
\end{table}

The headline values are
\begin{align}
r_\kappa(c{=}1)&=\infty,\\
r_\kappa(c{=}2)&\simeq 32\text{--}40\ \text{nodes}\quad (q=0.99).
\end{align}
At the crossover ($q=\tfrac12$) the value is more prior dependent, ranging from
$r_\kappa=14$ ($\alpha=3$) to $30$ ($\alpha=0.3$): a weaker prior demands a
larger network before communities emerge, while a stronger prior lowers the bar.
As the certainty requirement is tightened toward the almost-certain regime
of~\cite{vazquez2025ramsey}, this prior dependence washes out and $r_\kappa$
concentrates near $\sim35$ nodes.

Two features connect directly to~\cite{vazquez2025ramsey}. First, the ring is
built from a purely local rule with no node heterogeneity whatsoever, yet for
$c=2$ it acquires a finite Ramsey community number---an analytic confirmation
that community structure can be an emergent property of local connectivity
alone. Second, the contrast $r_\kappa(c{=}1)=\infty$ versus
$r_\kappa(c{=}2)<\infty$ shows that emergence is controlled by the interaction
range: a single-neighbour rule is too weak, but adding the second neighbour
suffices. Because the ring is deterministic, ``certainty'' here is the Bayesian
posterior weight of the partition rather than a frequency over random
realisations as in the ensembles of~\cite{vazquez2025ramsey}; the two notions
play the same role, namely the size at which the community description becomes
decisive.

Two caveats fix the scope. (i) The detection rule tested is the single balanced
bipartition against the null, so $r_\kappa(c{=}1)=\infty$ is a statement about
that comparison and does not preclude finer partitions being favoured under a
full multi-community scan. (ii) The crossing count $E_{12}=c(c+1)$ in
Eq.~\eqref{eq:E12} is constant only while the seams do not interact
($n_1,n_2>c$), which holds throughout the $c=2$ emergence regime, so that
conclusion is safe.

% ======================================================================
\section{Conclusions}
\label{sec:conclusions}
% ======================================================================

I have computed the Ramsey community number of the circulant ring lattice
analytically, using a Bernoulli stochastic block model with symmetric
$\mathrm{Beta}(\alpha,\alpha)$ priors as the community-detection rule. The
Bayesian evidence of the balanced two-community partition and of the
unpartitioned network are both single closed-form expressions,
Eqs.~\eqref{eq:Phalf}--\eqref{eq:Pzero}, and their ratio,
Eq.~\eqref{eq:R}, determines the transition exactly.

The main results are threefold. (1)~The plain cycle $C_n(1)$ is never
preferentially partitioned: its two-community posterior weight decays as
$n^{-(2\alpha+3)}$ and its Ramsey community number is infinite. (2)~The
next-nearest-neighbour ring $C_n(1,2)$ has a finite Ramsey community number,
$r_\kappa\simeq35$ nodes, above which the two-community description is preferred
with near-certainty; the log-evidence then grows as $(\ln 2)\,n$, independent of
the prior. (3)~The transition is governed by a single control parameter, the
interaction range $c$: a minimal amount of local connectivity is enough to make
the ring want to be broken.

These findings give an exactly solvable example of the community-emergence
phenomenon reported from simulations in~\cite{vazquez2025ramsey}, in a network
that is homogeneous, deterministic, and locally wired, and with no communities
built into its construction. They support the view that emergent communities are
a signature of local connectivity, and they make precise how the effect switches
on with interaction range. Natural extensions include the full partition profile
$P(n_1)$ over all $n_1$ to confirm that the balanced cut is the favoured one, the
scaling of $r_\kappa$ with larger $c$, and multi-community partitions of the ring,
each of which the present closed-form approach can address.

\begin{acknowledgments}
The calculations, computer scripts, and text of this manuscript were generated
by Claude Opus 4.8 (Anthropic). The analytical results were verified with an
accompanying open-source script at  \href{https://github.com/av2atgh/ring}{github.com/av2atgh/ring}.
\end{acknowledgments}

\bibliographystyle{apsrev4-2}
\bibliography{refs}

\end{document}